# Strain-Modulated Graphene Heterostructure as a Valleytronic Current Switch

Maverick Chauwin[1,2], Zhuo Bin Siu[1], and Mansoor Bin Abdul Jalil[1]

[1] Department of Electrical and Computer Engineering, National University of Singapore, Singapore

[2] Department of Physics, École polytechnique, 91128 Palaiseau, France

Correspondence and requests for materials should be addressed to M. B. A. J. (elembaj@nus.edu.sg).

## Abstract

Strain engineering is a promising approach for suppressing the OFF-state conductance in graphene-based devices that arises from Klein tunnelling. In this work, we derive a comprehensive tight-binding Hamiltonian for strained graphene that incorporates strain-induced effects that have been neglected hitherto, such as the distortion of the unit cell under strain, the effect of strain on the next-nearest neighbor coupling, and the second-order contributions of the strain tensor. We derive the corresponding low-energy effective Hamiltonian about the Dirac points and reformulate the boundary conditions at the interfaces between strained and unstrained graphene in light of additional terms in the Hamiltonian. By applying these boundary conditions, we evaluate the transmission across a strained graphene heterostructure consisting of a central segment sandwiched between two unstrained leads. Modulation of the transmitted current can be effected by varying the magnitude and direction of the applied strain, as well as by the applying a gate voltage. Based on realistic parameter values, we predict that high ON-OFF ratios of up to $10^{12}$ as well as high current valley polarization can be achieved in the strain-modulated device.

## Introduction

Graphene has attracted intensive research attention in the past years as a promising candidate for next-generation nanoelectronic devices [1 – 4]. The charge carriers in graphene are governed by the relativistic massless Dirac equation. As such, they experience the Klein tunnelling paradox [5 – 10] where a gate potential barrier becomes transparent to normally incident electrons, allowing 100% electron transmission through the barrier independently of its length and height. This property limits the applications of graphene-based devices because it hinders the suppression of the OFF-state conductance. One approach to address this issue is to confine the electrons by means of a localized magnetic field [11]. However, this in turn poses the new practical challenge of applying the required magnetic field strength at the scale of nanometers.

Strain engineering has emerged as a possible alternative [12 – 14]. The application of strain changes the electronic, thermal, optical, and chemical properties of the strained material [15 – 23], and affords two additional degrees of freedom, namely, the direction and the strength of the strain, for tuning these properties [24, 25]. Strain engineering is particularly applicable to two-dimensional materials like graphene, and there have been much efforts to modify the electronic transport properties of graphene via strain [24, 26 – 31]. Furthermore, it has been shown that a non-uniform strain could be modelled as an effective magnetic field [11, 32, 33] with an effective field strength of up to the order of 300 T [34]. An advantage of a strain-induced effective magnetic field as opposed to a real magnetic field is the valley asymmetry, which arises from the distortion of the lattice symmetry. This valley-dependent aspect can be utilized in potential valleytronic applications, in which the valley is considered as an additional internal degree of freedom [35 – 47]. Moreover, the high flexibility of graphene under mechanical pressure and novel state-of-the-art facilities and techniques in strain engineering [48 – 50] allow accurate control of the applied strains, which may reach 10% without destroying the graphene lattice [51]. Experimentally, one way of inducing strain is to deposit graphene onto a substrate so that a strain is induced by the difference between the crystal structures of the two. The strain can be further tuned by stretching the substrate [52 – 57]. Other methods of applying strain to graphene include the direct application of pressure on graphene monolayers by STM tips [58], or via gas inflation [59]. These techniques have the advantage of avoiding any graphene-substrate interactions which can introduce undesired modifications to the electronic properties of graphene.

In parallel to the aforementioned experimental studies, numerous theoretical valleytronics devices based on strained graphene have been proposed. Fogler *et. al.* were amongst the first to investigate the valley-resolved transmission in strained graphene heterojunctions [60]. They modelled the strain as a vector gauge-potential term in the Dirac fermion Hamiltonian due to the modulation of the hopping integrals between adjacent carbon atoms. A similar approach was taken in subsequent studies of valley-resolved transmission across various configurations of strained graphene segments sandwiched between unstrained leads [61 – 69].

Apart from the displacement of the Dirac point, the application of strain leads to additional changes. For instance, the modification of the real space unit cell due to the lattice distortion would need to be considered in addition to the modulation of the hopping integrals by the strain [70 – 72]. The breaking of the lattice symmetry in the presence of strain also leads to bandstructure features such as the displacement and opening of the Dirac points [73] and the distortion of the Fermi surfaces [74 – 78], which cannot be captured by the strain-gauge potential term. Although the strain-induced distortion of the Fermi surface has been considered in subsequent studies on the transmission in strained graphene heterojunctions [79 – 81], none of these works have considered all of the relevant effects of strain on the Fermi surface simultaneously, or the effect of the second-nearest neighbor coupling.

To address this gap, we derive a tight-binding Hamiltonian for strained graphene that incorporates the next-nearest coupling and the second-order contributions of the strain tensor while accounting for the distortion of the real space unit cell at the same time. We show that the previously neglected next-nearest coupling results in a tilt and a constant energy shift of the Dirac points. The tilt and shift have the important consequences of not only affecting the reciprocal space shape and size of the Fermi surface, which in turn affect the transport characteristics across an interface between a strained and unstrained segment, but also necessitates the derivation of a new set of appropriate boundary conditions for solving the transmission across the interface. The boundary conditions are then used to obtain analytic expressions for the transmission across a heterojunction consisting of a strained central segment with an applied gate potential sandwiched between two unstrained leads. Our numerical results show that a high ON-OFF current ratio can be attained by applying a gate voltage on the strained heterostructure. The current can also be modulated by changing either the direction or magnitude of the applied strain, and made highly valley-polarized.

## I.  Tight-binding Hamiltonian for Strained Graphene

We begin with preliminary discussions on the characteristics of unstrained graphene, i.e., its lattice structure, tight-binding Hamiltonian, dispersion relation, and positions of its Dirac points. This would form the framework for us to subsequently derive and analyze the corresponding properties of graphene under applied strain of arbitrary magnitude and direction.

### A.  Electronic properties of unstrained graphene

The unit cell of graphene is made of two atoms, which form two sublattices A and B when repeated along the following lattice vectors

$$\boldsymbol{L_1} = a\sqrt{3}(1,0), \boldsymbol{L_2} = a\sqrt{3}\left(\frac{1}{2}, \frac{\sqrt{3}}{2}\right), \text{ and } \boldsymbol{L_3} = \boldsymbol{L_2} - \boldsymbol{L_1} \tag{1}$$

where $a = 1.42$ Å [82] is the distance between two nearest-neighbors carbon atoms in unstrained graphene. The nearest-neighbor vectors from one sublattice to another are defined as:

$$\boldsymbol{d_1} = a(1,0), \boldsymbol{d_2} = a(-\sqrt{3}/2, -1/2), \text{ and } \boldsymbol{d_2} = a(\sqrt{3}/2, -1/2). \tag{2}$$

In the tight-binding approximation, the electrons hop from one atomic site to another. To simplify the problem, only the nearest- and next-nearest- neighbor hoppings are considered. In the second quantization formalism, the tight-binding Hamiltonian reads:

$$H = -t \sum_{\langle i,j \rangle, \sigma} \left(a_{\sigma,i}^{\dagger} b_{\sigma,j} + b_{\sigma,i}^{\dagger} a_{\sigma,j}\right) - t_n \sum_{\langle\langle i,j \rangle\rangle, \sigma} \left(a_{\sigma,i}^{\dagger} a_{\sigma,j} + b_{\sigma,i}^{\dagger} b_{\sigma,j}\right) \tag{3}$$

where $t = 2.9$ eV [83] is the nearest-neighbor hopping energy, and $t_n \approx 0.1t$ [84] is the next-nearest-neighbor hopping energy; $a_{\sigma,i}^{\dagger}$ ($a_{\sigma,j}$) annihilates (creates) an electron with spin $\sigma$ on the $i^{\text{th}}$ atom site in sublattice A and $b_{\sigma,i}^{\dagger}$ and $b_{\sigma,j}$ are similarly defined in sublattice B; $\langle i,j \rangle$ ($\langle\langle i,j \rangle\rangle$) is the set of integers such that sites $i$ and $j$ are nearest-neighbors (next-nearest-neighbours). Equation (3) can be written in reciprocal space using the Fourier transforms of the creation and annihilation operators. Defining $\Psi_{\boldsymbol{k},\sigma} = \left(a_{\boldsymbol{k},\sigma}, b_{\boldsymbol{k},\sigma}\right)^T$,

$$H(\boldsymbol{k}) = \sum_{\boldsymbol{k},\sigma} \Psi_{\boldsymbol{k},\sigma}^{\dagger} \underbrace{\begin{pmatrix} h'(\boldsymbol{k}) & h(\boldsymbol{k}) \\ h^*(\boldsymbol{k}) & h'(\boldsymbol{k}) \end{pmatrix}}_{\mathcal{H}(\boldsymbol{k})} \Psi_{\boldsymbol{k},\sigma}, \tag{4}$$

where

$$\begin{cases} h(\boldsymbol{k}) = -t \sum_{i=1}^{3} e^{-i\boldsymbol{k}\cdot\boldsymbol{d}_i} \\ h'(\boldsymbol{k}) = -2t_n \sum_{i=1}^{3} \cos(\boldsymbol{k}\cdot\boldsymbol{L}_i) \end{cases}. \tag{5}$$

The energy bands are the eigenvalues of $\mathcal{H}(\boldsymbol{k})$, and are given by

$$E_{\pm}(\boldsymbol{k}) = h'(\boldsymbol{k}) \pm |h(\boldsymbol{k})| = -t'f(\boldsymbol{k}) \pm \sqrt{3 + f(\boldsymbol{k})},$$
$$f(\boldsymbol{k}) = 2\cos(\sqrt{3}ak_x) + 4\cos\left(\frac{\sqrt{3}}{2}ak_x\right)\cos\left(\frac{3}{2}ak_y\right). \tag{6}$$

Unstrained graphene is gapless as the two energy bands touch at the Dirac points $\boldsymbol{K_D}$, which are defined through $h(\boldsymbol{K_D}) = \sum_{i=1}^{3} e^{-i\boldsymbol{K_D}\cdot\boldsymbol{d}_i} = 0$. Explicitly, the coordinates of the two Dirac points read as

$$a\boldsymbol{K_D^{\xi}} = \xi\left(\frac{4\pi}{3\sqrt{3}}; 0\right), \text{ where } \xi = \pm 1. \tag{7}$$

At low energies, the Hamiltonian around the Dirac points at $\boldsymbol{k^{\xi}} = \boldsymbol{K_D^{\xi}} + \boldsymbol{q}$, $\|a\boldsymbol{q}\| \ll 1$ is given by

$$\mathcal{H}(\boldsymbol{k^{\xi}}) = 3t_n\boldsymbol{\sigma^0} + v_F(\xi\hat{q}_x\boldsymbol{\sigma^x} - \hat{q}_y\boldsymbol{\sigma^y}) \tag{8}$$

where $\boldsymbol{\sigma^0}$ is the identity matrix, $\boldsymbol{\sigma^{x,y}}$ are the Pauli matrices, and $v_F = 3at/2$ is the Fermi velocity.

## B. Band structure of strained graphene

We now consider the application of a tensile or compressive strain of arbitrary magnitude and direction on the bond parameters and hopping integrals in graphene. Consider a graphene monolayer deformed by a unidirectional strain applied at an angle $\phi$ with the $x$-axis, which is described by the following strain tensor [85, 86]:

$$\boldsymbol{\epsilon} = \epsilon\begin{pmatrix} \cos^2\phi - \nu\sin^2\phi & (1+\nu)\cos\phi\sin\phi \\ (1+\nu)\cos\phi\sin\phi & \sin^2\phi - \nu\cos^2\phi \end{pmatrix} \tag{9}$$

where $\epsilon$ is the strain strength, which is positive for tensile strains and negative for compressive strains, and $\nu = 0.149$ is the Poisson's ratio [87]. The lattice vectors are modified to $\boldsymbol{L}_i = \boldsymbol{L}_i^{(0)} \cdot$ $(I_2 + \epsilon)$ and $d_i = d_i^{(0)} \cdot (I_2 + \epsilon)$, where $L_i^{(0)}$ and $d_i^{(0)}$ are the undeformed lattice vectors, and $I_2$ is the identity matrix.

The strain affects the graphene monolayer in two significant ways: i) the hexagonal symmetry is broken, and ii) the hopping energies to the nearest and next-nearest neighbors are no longer isotropic but depend on the direction of the hop. The change in the hopping integrals with bond length can be modelled as a decreasing exponential [88,89]:

$$t_i = t \exp\left(-\beta(|d_i|/a - 1)\right)$$
$$t_{ni} = t \exp\left(-\beta(|L_i|/a - 1)\right) = t_n \exp\left(-\beta(|L_i|/a - \sqrt{3})\right), \tag{10}$$

where the Grüneisen parameter $\beta$ was set to 3.37 [90] to satisfy the identity $t_n = t \exp\left(-\beta(\sqrt{3} - 1)\right)$.

While the Hamiltonian $\mathcal{H}$ for strained graphene has a similar form to that for unstrained graphene Eq. (4), i.e.,

$$\begin{cases} h(\boldsymbol{k}) = -\displaystyle\sum_{i=1}^{3} t_i e^{-i\boldsymbol{k} \cdot \boldsymbol{d_i}} \\ h'(\boldsymbol{k}) = -2\displaystyle\sum_{i=1}^{3} t_{ni} \cos(\boldsymbol{k} \cdot \boldsymbol{L_i}) \end{cases}, \tag{11}$$

the hopping integrals in Eq. (11) assume different values between different pairs of neighbors because of the broken symmetry in strained graphene. By comparison, in the case of unstrained graphene described by Eq. (4), the hopping integrals $t_i$ and $t_{ni}$ have the same values between all the three nearest and six next-nearest neighboring atoms, respectively.

## C. Dirac point coordinates

Substituting the deformed lattice vectors and hopping integrals of strained graphene into Eq. (11), and expanding the Taylor series to the second order in the small strain parameter $\epsilon$ for $|\boldsymbol{d_i}|, |\boldsymbol{L_i}|, t_i$, and $t_{ni}$, the Dirac point coordinates to the quadratic approximation in strain can be derived [91, 92]. Assuming $\boldsymbol{K_\epsilon^\xi} = \boldsymbol{K_D^\xi} + \boldsymbol{K_1^\xi}\epsilon + \boldsymbol{K_2^\xi}\epsilon^2 = \boldsymbol{K_D^\xi} + \boldsymbol{\Delta K_\epsilon^\xi}$, the identity $\sum_{i=1}^3 t_i e^{-i\boldsymbol{K_\epsilon^\xi}\cdot\boldsymbol{d_i}} = 0$ can be expanded and, then, by term-wise comparison, we obtain the analytic expressions for the coefficients $\boldsymbol{K_1^\xi}$ and $\boldsymbol{K_2^\xi}$ for the strain-induced displacement of the Dirac points in terms of the strain parameters $\phi$ and $\nu$:

$$aK_{1x}^\xi = \xi\left[\frac{4\pi}{3\sqrt{3}}(\nu\sin^2\phi - \cos^2\phi) - \frac{\beta(1+\nu)}{2}(\sin^2\phi - \cos^2\phi)\right],$$

$$aK_{1y}^\xi = -\xi\left(\frac{4\pi}{3\sqrt{3}} + \beta\right)(1+\nu)\cos\phi\sin\phi\,,$$

$$aK_{2x}^\xi = \xi\left[\frac{4\pi}{3\sqrt{3}}(\cos^4\phi + (1+\nu^2)\cos^2\phi\sin^2\phi + \nu^2\sin^4\phi)\right.$$

$$-\frac{\beta(4\beta+1)(1+\nu)^2}{16}(\cos^4\phi - 6\cos^2\phi\sin^2\phi + \sin^4\phi) \qquad (12)$$

$$\left. -\frac{\beta(1+\nu)}{2}(\cos^4\phi - 3(1+\nu)\cos^2\phi\sin^2\phi + \nu\sin^4\phi)\right],$$

$$aK_{2y}^\xi = \xi(1+\nu)\cos\phi\sin\phi\left[\frac{4\pi}{3\sqrt{3}}(1-\nu)(\cos^2\phi + \sin^2\phi)\right.$$

$$+\frac{\beta(4\beta+1)(1+\nu)}{4}(\cos^2\phi - \sin^2\phi)$$

$$\left. +\frac{\beta}{2}(-(1+3\nu)\cos^2\phi + (3+\nu)\sin^2\phi)\right].$$

As can be seen in the above, the Dirac points are displaced in momentum space in both the $k_x$ and $k_y$ directions. Of the two, the displacement in the transverse direction, i.e., in the $k_y$ direction, would be more crucial in terms of its effects on the transport characteristics (as we shall see later). The displacement of the Dirac points along the $k_y$-axis vanishes whenever $\phi \equiv 0\ [\pi/2]$, i.e., when the strain is applied in the zigzag or the armchair direction. Moreover, to the first order in strain, the shift along the $k_y$ direction is maximized when $\phi = \pi/4$.

## II. Effective Hamiltonian in the low-energy approximation

For practical applications, it is useful to derive a low-energy Hamiltonian close to the Dirac points. In the vicinity of the Dirac points, the electron wave vectors can be written in the form $a\mathbf{k} = a\mathbf{K}_\epsilon^\xi + a\mathbf{q}$, where $\|a\mathbf{q}\| \ll 1$. Note that the effective Hamiltonian has to be expanded around the displaced Dirac points of strained graphene rather than the original Dirac points of unstrained graphene, in order to obtain the correct Fermi velocities [91, 92]. Expanding the effective Hamiltonian around the displaced Dirac points, the following general form is obtained after straightforward algebra:

$$\mathcal{H}(\xi \mathbf{K}_\epsilon + \mathbf{q}) = v_F \big[ w\boldsymbol{\sigma^0} + \xi t_n/t \left( w_{0x}\hat{q}_x + w_{0y}\hat{q}_y \right)\boldsymbol{\sigma^0} \\ + \xi \left( w_{1x}\hat{q}_x + w_{1y}\hat{q}_y \right)\boldsymbol{\sigma^x} - \left( w_{2x}\hat{q}_x + w_{2y}\hat{q}_y \right)\boldsymbol{\sigma^y} \big]$$

(13)

where $\boldsymbol{\sigma^0}$ is the $2 \times 2$ identity matrix, $\boldsymbol{\sigma^{x,y}}$ are the Pauli matrices, and $v_F$ is the Fermi velocity. The various $w_{ij}$ coefficients in the above equation are all functions of the strain parameters. These are rather involved, and are explicitly given by:

$$w = \frac{2t_n}{at} \times \left[1 + \frac{\beta\sqrt{3}}{2}(\nu-1)\epsilon + \frac{3\beta^2}{8}(\nu-1)^2\epsilon^2 + \frac{\beta^2\sqrt{3}}{8}(1+\nu)^2\epsilon^2 - \frac{\beta^2\sqrt{3}}{16}(1+\nu)^2\epsilon^2\right],$$

$$w_{0x} = \frac{3\beta(\sqrt{3}-1)}{2}(1+\nu)(\cos^2\phi - \sin^2\phi)\epsilon + \frac{3\beta^2\sqrt{3}}{8}(1+\nu)[(3-\nu)\cos^4\phi - 6(1+\nu)\cos^2\phi\sin^2\phi + (3\nu-1)\sin^4\phi]\epsilon^2$$

$$+ \frac{3\beta^2}{2}(1+\nu)[(\nu-2)\cos^4\phi + 3(1+\nu)\cos^2\phi\sin^2\phi + (1-2\nu)\sin^4\phi]\epsilon^2 + \frac{3\beta\sqrt{3}}{16}(1+\nu)[(7-\nu)\cos^4\phi - 18(1+\nu)\cos^2\phi\sin^2\phi$$

$$+ (7\nu-1)\sin^4\phi]\epsilon^2 + \frac{3\beta}{16}(1+\nu)[-(9+\nu)\cos^4\phi + 30(1+\nu)\cos^2\phi\sin^2\phi - (9\nu+1)\sin^4\phi]\epsilon^2,$$

$$w_{0y} = 3\beta(1-\sqrt{3})(1+\nu)\cos\phi\sin\phi\,\epsilon + \beta^2\left[3\sqrt{3}(\nu\cos^2\phi - \sin^2\phi) + \frac{3}{2}\big((1-5\nu)\cos^2\phi + (5-\nu)\sin^2\phi\big)\right](1+\nu)\cos\phi\sin\phi\,\epsilon^2$$

$$+ \beta\left[\frac{3\sqrt{3}}{4}\big((5\nu+1)\cos^2\phi - (\nu+5)\sin^2\phi\big) + \frac{3}{4}[-(7\nu+3)\cos^2\phi + (3\nu+7)\sin^2\phi]\right](1+\nu)\cos\phi\sin\phi\,\epsilon^2,$$

$$w_{1x} = 1 + (1-\beta)(\cos^2\phi - \nu\sin^2\phi)\epsilon + \frac{\beta^2}{4}[(-\nu^2 - 2\nu + 1)\cos^4\phi + 4(\nu^2+\nu+1)\cos^2\phi\sin^2\phi + (\nu^2-2\nu-1)\sin^4\phi]\epsilon^2$$

$$+ \frac{\beta}{8}[(-\nu^2 - 2\nu - 9)\cos^4\phi + (-6\nu^2 + 4\nu - 6)\cos^2\phi\sin^2\phi + (-9\nu^2 - 2\nu - 1)\sin^4\phi]\epsilon^2,$$

$$w_{2x} = w_{1y} = (1-\beta)(1+\nu)\cos\phi\sin\phi\,\epsilon + \left[\beta^2(\cos^2\phi - \nu\sin^2\phi) + \frac{\beta}{4}\big((5\nu-3)\cos^2\phi + (3\nu-5)\sin^2\phi\big)\right](1+\nu)\cos\phi\sin\phi\,\epsilon^2$$

$$w_{2y} = 1 + (1-\beta)(-\nu\cos^2\phi + \sin^2\phi)\epsilon + \frac{\beta^2}{2}[\nu^2\cos^4\phi - (\nu^2+4\nu+1)\cos^2\phi\sin^2\phi + \sin^4\phi]\epsilon^2$$

$$+ \frac{\beta}{2}[-2\nu^2\cos^4\phi - (3\nu^2 + 2\nu + 3)\cos^2\phi\sin^2\phi - 2\sin^4\phi]\epsilon^2. \tag{14}$$

Equations (13) and (14) define the low-energy Hamiltonian of graphene that captures the full effects of strain up to the second order in the strain parameter. This low-energy Hamiltonian constitutes the main result of the paper and provides a more complete description of the effect of strain on the graphene low-energy dispersion compared to earlier treatments [92], most of which have modelled the effects of strain as a single gauge potential [60 – 69]. As mentioned earlier, the strain gauge potential would only capture the shifting of the Dirac points up to some extent. However, it does not capture other strain-induced changes, such as the distortion of the Fermi surface from the circular shape, as well as the tilting of the Dirac cones, which are described by the effective Hamiltonian in Eq. (13). We summarize the major strain-induced physics in the vicinity of the Dirac points beyond the strain gauge potential treatment below:

- Because the low energy Hamiltonian is expanded around the strained Dirac points $K_\epsilon^\xi$, the Fermi surface is shifted by $\Delta K_\epsilon^\xi$, and centred around the strained Dirac points.

- A consequence of the breaking of the honeycomb lattice symmetry is the inequality $|w_{1x}q_x + w_{1y}q_y| \neq |w_{2x}q_x + w_{2y}q_y|$ in Eq. 13. The Fermi surface is no longer circular as for unstrained graphene but is distorted to form an ellipse.

- Because $(w_{1x}\hat{q}_x + w_{1y}\hat{q}_y)\boldsymbol{\sigma^x}$ has a $\hat{q}_y$ dependence and $(w_{2x}\hat{q}_x + w_{2y}\hat{q}_y)\boldsymbol{\sigma^y}$ a $\hat{q}_x$ dependence, the elliptical Fermi surface is rotated by the strain, i.e., its minor and major axes are no longer coincident with the $k_x$ and $k_y$ axes. This rotation is valley-dependent.

- $w\boldsymbol{\sigma^0}$ corresponds to a global shift in energy due to the strain, akin to an additional applied electrical potential.

- $\xi t_n/t \left(w_{0x}\hat{q}_x + w_{0y}\hat{q}_y\right)\boldsymbol{\sigma^0}$ implies that the Dirac cones are tilted in opposite directions in the two valleys. This term vanishes when the strength $\epsilon$ is equal to zero, and originates from the next-nearest-neighbor hoppings, as can be seen from its proportionality to $t_n$.

In particular, the last two properties emerge only with the inclusion of the next-nearest-neighbor coupling, which has been neglected in previous works [89].

## A. *Dispersion relation and strained Fermi surface*

From the low-energy Hamiltonian Eq. (13), the energy eigenvalues can be straightforwardly derived as

$$E_{\pm} = v_F \left[ w + \xi \frac{t_n}{t} \left(w_{0x}q_x + w_{0y}q_y\right) \right.$$
$$\left. \pm \sqrt{\left(w_{1x}q_x + w_{1y}q_y\right)^2 + \left(w_{2x}q_x + w_{2y}q_y\right)^2} \right]. \tag{15}$$

From Eq. (15), we can see that the Fermi surface constitutes a solution of the implicit equation $Aq_x^2 + Bq_xq_y + Cq_y^2 + Pq_x + Qq_y + R = 0$, which is the equation of a conic. The coefficients of the conic equation are as follows:

$$A = w_{1x}^2 + w_{2x}^2 - \left(\frac{t_n}{t} w_{0x}\right)^2,$$

$$B = 2\left(w_{1x}w_{1y} + w_{2x}w_{2y} - \left(\frac{t_n}{t}\right)^2 w_{0x}w_{0y}\right),$$

$$C = w_{1y}^2 + w_{2y}^2 - \left(\frac{t_n}{t} w_{0y}\right)^2,$$

$$P = \xi \frac{2t_n}{t} w_{0x} \left(\frac{E}{v_F} - w\right), \tag{16}$$

$$Q = \xi \frac{2t_n}{t} w_{0y} \left(\frac{E}{v_F} - w\right),$$

$$R = -\left(\frac{E}{v_F} - w\right)^2.$$

The sign of the discriminant $\Delta = 4AC - B^2$ characterizes the conic type of the Fermi surface: The Fermi surface is elliptical, parabolic, or hyperbolic when $\Delta$ is positive, zero, or negative, respectively. Here, we focus only on the elliptical case. Figure 1 shows the contour plots of the conduction bands calculated with Eq. (15) for various strain configurations. These plots reveal the distortion of the contour surfaces from the circular shape in unstrained graphene (Fig. 1a) to the elliptical shape in graphene strained along either the armchair or zigzag directions (Figs. 1b and 1c) and to the rotated elliptical shape in graphene strained along some arbitrary direction (Figs. 1d to 1f).

To further analyze the shape of the Fermi surfaces, we consider the principal axes of the conic. The axes are rotated with respect to the $q_x$-axis at an angle $\theta_r \equiv \frac{1}{2}\arctan\left(\frac{B}{A-C}\right)\left[\frac{\pi}{2}\right]$. Introducing the constants

$$A' = A\cos^2\theta_r + B\cos\theta_r\sin\theta_r + C\sin^2\theta_r,$$
$$C' = A\sin^2\theta_r - B\cos\theta_r\sin\theta_r + C\cos^2\theta_r,$$
$$P' = P\cos\theta_r + Q\sin\theta_r,$$
$$Q' = -P\sin\theta_r + Q\cos\theta_r \qquad (17)$$

the lengths of the principal axes read

$$\mathrm{PA}_1 = \sqrt{\frac{P'^2}{4A'^3} + \frac{Q'^2}{4C'^2A'} - \frac{R}{A'}},$$

$$\mathrm{PA}_2 = \sqrt{\frac{P'^2}{4A'^2C'} + \frac{Q'^2}{4C'^3} - \frac{R}{C'}}, \qquad (18)$$

and the ellipse is centered around

$$\left(\frac{4\pi}{3a\sqrt{3}} + \Delta K_{\epsilon x}^{\xi} - \frac{P'\cos\theta_r}{2A'} + \frac{Q'\sin\theta_r}{2C'} \; ; \Delta K_{\epsilon y}^{\xi} - \frac{P'\sin\theta_r}{2A'} - \frac{Q'\cos\theta_r}{2C'}\right). \qquad (19)$$

One can see that the shifting of the center of the ellipse stems from two different effects, namely, the displacement of the Dirac points $\boldsymbol{\Delta K_{\epsilon}^{\xi}}$ and an additional shift from the tilting of the Dirac cones due to finite $w_{0x}$ and $w_{0y}$.

## III.   Transmission in a strained graphene heterojunction

### 1) Derivation of the transmission coefficient

Having derived the effective Hamiltonian for strained graphene in the previous section, we will now apply it to model the effects of strain on the transport properties across the graphene heterojunction system shown in Fig. 2. Here, we consider a graphene heterojunction in which a localized strain is applied to the central region (Region II) in addition to the gate potential. To achieve this localized strain, the metallic contact used to apply the gate potential may also be made to serve the additional function of controlling the strain in the central part of the device. Previous works have shown that application of a mechanical force on the metallic gate may impact the local strain of the graphene [52 – 57]. The central segment is sandwiched between unstrained segments connected to a source (Region I) and drain (Region III) leads, as depicted in Fig. 2. We assume that electrons enter the system from the source with the energy $E_0$ and the incident angle $\varphi$. The system is invariant in the $y$-direction, and thus the momentum $q_y$ is conserved throughout the device.

**Unstrained regions I ($x < 0$) and III ($x > L$)**

In these regions, the low-energy effective Hamiltonian expanded around the $\xi$ valley reads:

$$\mathcal{H}^{I/III} = 3t_n \boldsymbol{\sigma^0} + v_F \left( \xi \hat{q}_x \boldsymbol{\sigma^x} - \hat{q}_y \boldsymbol{\sigma^y} \right). \tag{20}$$

The electron wave functions can be written as

$$|\psi_I\rangle = \frac{1}{\sqrt{2}} \begin{pmatrix} 1 \\ s\xi e^{-i\xi\varphi} \end{pmatrix} e^{iq_x x} + \frac{r}{\sqrt{2}} \begin{pmatrix} 1 \\ -s\xi e^{i\xi\varphi} \end{pmatrix} e^{-iq_x x},$$

$$|\psi_{III}\rangle = \frac{t}{\sqrt{2}} \begin{pmatrix} 1 \\ s\xi e^{-i\xi\varphi} \end{pmatrix} e^{iq_x x}, \tag{21}$$

where the effective energy is defined as

$$\mathcal{E} = \frac{E_0 - 3t_n}{v_F} \quad \text{and} \quad s = \text{sign}(\mathcal{E}), \tag{22}$$

and the wave vector reads

$$q_x = q\cos\varphi, \qquad q_y = q\sin\varphi,$$

$$q = \sqrt{q_x^2 + q_y^2} = |\mathcal{E}|. \tag{23}$$

The probability for the electron to be reflected by the barrier (respectively to pass through the barrier) is $|r|^2$ (respectively $|t|^2$). There is no discontinuity in region III that could reflect the electron backwards, and thus the state with a negative wavevector is omitted in $|\psi_{III}\rangle$.

**Strained region II ($0 < x < L$)**

Here, a gate potential $V_0$ is applied to the strained region. The low-energy effective Hamiltonian with respect to the *unstrained $\boldsymbol{K^\xi}$* Dirac point can be derived from Eq. (13):

$$\mathcal{H}^{II}(\xi \mathbf{K} + \mathbf{q}) = V_0 \boldsymbol{\sigma^0} + v_F \big[ w \boldsymbol{\sigma^0} + \xi t_n/t \left( q_0 + w_{0x}\hat{q}_x + w_{0y}\hat{q}_y \right) \boldsymbol{\sigma^0}$$
$$+ \xi \left( q_1 + w_{1x}\hat{q}_x + w_{1y}\hat{q}_y \right) \boldsymbol{\sigma^x} - \left( q_2 + w_{2x}\hat{q}_x + w_{2y}\hat{q}_y \right) \boldsymbol{\sigma^y} \big] \tag{24}$$

where

$$q_i = -\left( w_{ix}\Delta K_{\epsilon x}^{\xi} + w_{iy}\Delta K_{\epsilon y}^{\xi} \right). \tag{25}$$

The conservation of the energy $E_0$ and transverse momentum $q_y$ leads to the allowed values for the momentum $q_x$ in the central region taking the form of

$$q_x^{\pm} = \frac{-b_x \pm \sqrt{b_x^2 - 4a_x c_x}}{2a_x} \tag{26}$$

where

$$a_x = w_{1x}^2 + w_{2x}^2 - \left( \frac{t_n}{t} \right)^2 w_{0x}^2$$
$$b_x = 2w_{1x}(q_1 + w_{1y}q_y) + 2w_{2x}(q_2 + w_{2y}q_y)$$
$$-2w_{0x}\frac{t_n}{t}\left( \xi w + \frac{t_n}{t}(q_0 + w_{0y}q_y) - \xi \frac{E_0 - V_0}{v_F} \right)$$
$$c_x = \left( v_F q_1 + w_{1y}q_y \right)^2 + \left( v_F q_2 + w_{2y}q_y \right)^2 \tag{27}$$
$$-\left( \frac{E_0 - V_0}{v_F} + \frac{t_n}{t}(q_0 + q_y w_{0y}) + w \right)^2$$
$$+(1 - v_F^2)(q_1^2 + q_2^2) + 2q_y(1 - v_F)(q_1 w_{1y} + q_2 w_{2y})$$
$$+\frac{4(E_0 - V_0)wt + 2 t_n(q_0 + q_y w_{0y})(-v_F w(\xi - 1) + (E_0 - V_0)(1 + \xi)}{t\, v_F}.$$

By defining the effective wave vectors

$$Q_0^{\pm} = w + \xi \frac{t_n}{t}(q_0 + w_{0x}q_x^{\pm} + w_{0y}q_y),$$
$$Q_x^{\pm} = \xi(q_1 + w_{1x}q_x^{\pm} + w_{1y}q_y),$$
$$Q_y^{\pm} = -(q_2 + w_{2x}q_x^{\pm} + w_{2y}q_y), \tag{28}$$
$$Q^{\pm} = \sqrt{\left( Q_x^{\pm} \right)^2 + \left( Q_y^{\pm} \right)^2} = s_{\pm}\left( \frac{E_0 - V_0}{v_F} - Q_0^{\pm} \right),$$

the wave function of the electron in region II can then be expressed as

$$|\psi_{II}\rangle = \frac{a}{Q^{+}\sqrt{2}}\begin{pmatrix} Q^{+} \\ s_{+}(Q_x^{+} + iQ_y^{+}) \end{pmatrix} e^{iq_x^{+}x} + \frac{b}{Q^{-}\sqrt{2}}\begin{pmatrix} Q^{-} \\ s_{-}(Q_x^{-} + iQ_y^{-}) \end{pmatrix} e^{iq_x^{-}x}. \tag{29}$$

**Boundary conditions**

Having defined the electron wavefunctions in the three regions of the strained graphene heterostructure, we now apply the boundary conditions linking these wavefunctions across the interfaces. In most quantum mechanical systems, one of the basic boundary conditions is the continuity of the wave function across any interface. However, in the present case, there is a fundamental discontinuity in the Hamiltonian across the interfaces which results in the discontinuity of the wave function [93]. Discontinuous wavefunctions have previously been discussed in Dirac fermion heterojunctions where the Fermi velocity differs across the heterojunction interface [94 – 96], and the more fundamental conserved quantity is the particle current or flux. However, in these previous works, the velocity operators across the interfaces differ only by a multiplicative factor. In other words, denoting the Hamiltonians of the two regions on either side of the interface as $\mathcal{H}^a$ and $\mathcal{H}^b$, we have the relation

$$\partial_{q_x}\mathcal{H}^a = \nu\partial_{q_x}\mathcal{H}^b \tag{30}$$

where $\nu$ is a scalar. In this case, current conservation can be ensured by imposing the boundary condition $\sqrt{\nu}\psi^a(0) = \psi^b(0)$ at the interface. It is, however, evident from Eqs. (20) and (24) that Eq. (30) will not be satisfied at the interfaces in our present system between the strained and unstrained segments, since $\partial_{q_x}\mathcal{H}^{II}$ in the strained segment contains terms proportional to $\sigma^0$ and $\sigma^y$ due to the applied strain, which are absent in $\partial_{q_x}\mathcal{H}^{I/III}$. This requires a more refined set of boundary conditions in order to preserve the conservation of current.

We assume that the discontinuity in the wavefunction satisfies the general linear relations

$$\psi_{\mathrm{I}}(x = 0) = \boldsymbol{M}\psi_{\mathrm{II}}(x = 0),$$
$$\psi_{\mathrm{III}}(x = L) = \boldsymbol{M}\psi_{\mathrm{II}}(x = L), \tag{31}$$

where the discontinuity matrix $\boldsymbol{M}$ satisfies the following:

$$\boldsymbol{M}^\dagger \boldsymbol{\sigma^x} \boldsymbol{M} = \frac{w_{0x} t_n}{t} \boldsymbol{\sigma^0} + w_{1x} \boldsymbol{\sigma^x} - \xi w_{2x} \boldsymbol{\sigma^y},$$
$$\lim_{\epsilon \to 0} \boldsymbol{M} = \boldsymbol{\sigma^0}, \tag{32}$$

in order to ensure current conservation. When the strain $\epsilon$ tends towards zero, the discontinuity in the Hamiltonian vanishes, and so the wave function must become continuous across the boundaries, i.e., the discontinuity matrix $\boldsymbol{M}$ tends towards the identity matrix in the limit $\epsilon \to 0$.

We now set out to determine the components of the discontinuity matrix $\boldsymbol{M}$. The time-invariance of the system along the $y$-axis results in the conservation of particle flux $\partial_t \rho$, where $\rho = |\psi|^2$ is the electron probability. Differentiating the Schrödinger equation $i\hbar \partial_t \psi = \widehat{H} \psi$ twice with respect to $x$, we obtain the continuity equation

$$\frac{\partial \rho}{\partial t} + \boldsymbol{\nabla} \cdot \boldsymbol{j} = 0 \tag{34}$$

where the probability current is

$$\boldsymbol{j} = \begin{cases} -\dfrac{v_F}{\hbar} \begin{pmatrix} \langle \psi | \xi \boldsymbol{\sigma^x} | \psi \rangle \\ \langle \psi | -\boldsymbol{\sigma^y} | \psi \rangle \end{pmatrix} & \text{, in regions I/III} \\ -\dfrac{v_F}{\hbar} \begin{pmatrix} \left\langle \psi \middle| \dfrac{\xi w_{0x} t_n}{t} \boldsymbol{\sigma^0} + \xi w_{1x} \boldsymbol{\sigma^x} - w_{2x} \boldsymbol{\sigma^y} \middle| \psi \right\rangle \\ \left\langle \psi \middle| \dfrac{\xi w_{0y} t_n}{t} \boldsymbol{\sigma^0} + \xi w_{1y} \boldsymbol{\sigma^x} - w_{2y} \boldsymbol{\sigma^y} \middle| \psi \right\rangle \end{pmatrix} & \text{, in region II} \end{cases} \tag{35}$$

The conservation of the particle flux in the $x$-direction leads to the following current conservation equations:

$$\boldsymbol{j_I}(x = 0) \cdot \boldsymbol{e_x} = \boldsymbol{j_{II}}(x = 0) \cdot \boldsymbol{e_x},$$
$$\boldsymbol{j_{III}}(x = L) \cdot \boldsymbol{e_x} = \boldsymbol{j_{II}}(x = L) \cdot \boldsymbol{e_x}. \tag{36}$$

These conditions imply the identity $1 = |r|^2 + |t|^2$: the incoming wave splits into a reflected and a transmitted wave with no absorption by the strained material. However, Eqs. (32) and (36) are not sufficient to define a unique solution for the matrix $\boldsymbol{M}$. To obtain a unique solution, it is reasonable to impose a symmetric form for $\boldsymbol{M}$ assuming symmetry in the discontinuity at the interface for the two sublattices. This yields

$$\boldsymbol{M} = \begin{pmatrix} z & x_0 \\ x_0 & z^* \end{pmatrix} \tag{37}$$

where

$$z = \sqrt{\frac{w_{1x}}{2}\left(1 + \sqrt{1 + \frac{w_{2x}^2 - (t_n/t)^2 w_{0x}^2}{w_{1x}^2}}\right)} + i\left(-\xi\frac{w_{2x}}{2\mathfrak{Re}(z)}\right),$$
$$x_0 = \frac{t_n}{t} \times \frac{w_{0x}}{2\mathfrak{Re}(z)}. \tag{38}$$

Note that the chosen symmetric form of $\boldsymbol{M}$ is such that the two diagonal elements have the same magnitude while the off-diagonal terms have the same values when the next-nearest-neighbor coupling vanishes (zero-tilt scenario).

Defining

$$\omega_+^{\pm} = +s\xi e^{-i\xi\varphi}\left(z + s_{\pm}x_0\frac{Q_x^{\pm} + iQ_y^{\pm}}{Q^{\pm}}\right) - \left(x_0 + s_{\pm}z^*\frac{Q_x^{\pm} + iQ_y^{\pm}}{Q^{\pm}}\right),$$
$$\omega_+^{\pm} = -s\xi e^{+i\xi\varphi}\left(z + s_{\pm}x_0\frac{Q_x^{\pm} + iQ_y^{\pm}}{Q^{\pm}}\right) - \left(x_0 + s_{\pm}z^*\frac{Q_x^{\pm} + iQ_y^{\pm}}{Q^{\pm}}\right), \tag{39}$$

the transmission coefficient $T = |t|^2$ can then be expressed as:

$$T = \frac{4q_x^2}{q^2} \times \left| s_+\frac{Q_x^+ + iQ_y^+}{Q^+} - s_-\frac{Q_x^- + iQ_y^-}{Q^-} \right|^2 \times \left| \frac{(|z|^2 - x_0^2)e^{i(q_x^+ + q_x^- - q_x)L}}{\omega_-^{\pm}\omega_+^- e^{iq_x^- L} - \omega_+^{\pm}\omega_-^- e^{iq_x^+ L}} \right|^2. \tag{40}$$

The above analytic expression for the transmission across the strained graphene heterostructure constitutes another key result of this work.

## *Results and discussion*

In this section we present three exemplary sets of numerical calculations based on our analytical derivation of the low-energy Hamiltonian of strained graphene, and the corresponding transmission across the strained-unstrained graphene heterojunctions shown in Fig. 2. We consider the transmission for the two valleys under varying electrostatic potential and uniaxial strain in the central segment.

Figure 3 shows the transmission coefficients in the two valleys as functions of the incidence angle $\varphi$, and the Fermi surfaces (FSs) in the leads (grey) and in the strained central segment (red for $K$ valley, blue for $K'$ valley) under varying gate potential $V_0$ applied to the central segment. In the heterojunction device, the transverse momentum $q_y$ is conserved across the interfaces. This implies that the range of $q_y$ spanned by the Fermi surface of the leads should overlap with the $q_y$ range of the propagating states (i.e., states with real values of $q_x$) in the strained central segment for significant transmission to occur. In the absence of any overlap, the transmission would be suppressed since only evanescent states with complex or imaginary values of $q_x$ are present in the central segment. Therefore, in Fig. 3a, we plot the projections of the dispersion relations of the unstrained leads and the strained central segment on the $E$-$q_y$ plane. The range of overlapping $q_y$ between the lead and central segment FSs can be modulated by shifting the dispersion relation of the central segment along the energy axis by the application of the gate potential $V_0$. Four energy levels are depicted in the central segment corresponding to the four values of $V_0 = 0, \pm 450 \text{ meV}, 800 \text{ meV}$, as shown in Fig. 3a. The equal-energy contours at these energies can be compared to that of the source at the Fermi energy $E_f = 100$ meV in order to determine the extent of overlap between the Fermi surfaces of the source and central segments.

It can be seen from Fig. 3b and Fig. 3c that the strain in the central segment leads to the displacement of its Dirac point in $k$-space relative to the lead Dirac point along opposite directions for the two valleys. However, the displacement along the energy axis occurs in the same direction for both valleys. These strain-induced displacements of the central segment Dirac point result in an absence of overlap between the $q_y$ values spanned by the lead and the central segment FSs at zero or small values of $|V_0|$, as shown in Fig. 3b for the exemplary value of $V_0 = 0$. There are therefore no propagating states in the central segments for those values of $q_y$ spanned by the lead states. This results in negligible transmission due to the exponential decay of the evanescent states in the central segment away from the source lead.

As $V_0$ is increased to larger *positive* values, the FS of the central segment shifts lower into the hole-like states with a corresponding increase in the $k$-space diameter of the FS. Hence, the range of $q_y$ spanned by the central segment states begins to overlap with those of the lead states. A partial overlap is depicted in Fig. 3c corresponding to $V_0 = 450$ meV. There is significant transmission at positive (negative) incidence angles for the $K'$ $(K)$ valley because these incidence angles (which are marked out on the lead FSs in the lower inset of Fig. 3c, and shaded red for the $K'$ valley in the transmission plot of Fig. 3c) correspond to the overlapping range of $q_y$ spanned by the FSs of the source and the central segment. As $V_0$ is further increased, the central segment FS grows in extent and eventually encompasses the entire FS of the source lead, as depicted in Fig. 3d for $V_0 = 800$ meV. In this case, the $q_y$ values spanned by the entire source FS have corresponding propagating states in the central segment FS, i.e., transmission occurs for all incidence angles.

In addition to the extent of overlap between the FSs of the source and central segment (and hence the presence of forward propagating states in the central segment), a secondary factor that determines the transmission probability is the relative orientations of the pseudospins of the propagating eigenstates in the source and the central segment. In the lower right corner of Fig. 3d, we plot the orientations of the pseudospin in the lead and the central segment for two values of $q_y$ in the FSs of the source and central segment at the $K$ valley. (Note that because the central segment states are hole-like states, the states that propagate in the positive $x$-direction lie on the left half of the FS whereas the source states are particle-like states and the corresponding propagating states lie on the right half of the FS). In the $K$ valley, the pseudospin directions in the source and central segment states are more closely aligned at negative values of incidence angles (the green pair of arrows) compared to positive values of incidence angles (the black pair of arrows). This results in the preferential transmission of the $K$ valley at negative incidence angles, while the reverse occurs in the $K'$ valley where positive incidence angles are preferred. These trends are in agreement with the angular plots on the left of Fig. 3d.

This situation is reversed when $V_0$ has a large negative value so that both the enclosing FS of the central segment and the enclosed FS of the lead have *particle*-like states, as shown in Fig. 3e for $V_0 = -450$ meV. In this case, the forward propagating states in both the source and central segment lie on the right half of their respective FSs. At the $K$ valley, their corresponding pseudospin orientations are more closely aligned at positive incidence angles (black arrows) than at negative incidence angles (green arrows). The reverse occurs for the $K'$ valley, resulting in the preferential transmission at negative (positive) angles for the $K$ ($K'$) valley.

Besides varying the gate potential in the strained segment, the transmission profile can also be modulated by changing the magnitude of the strain. Figure 4 shows the transmission profile and the FSs of the lead and central segments in the two valleys when a fixed gate potential of $V_0 = 450$ meV is applied to the central segment and the strain application angle of $\phi = \frac{\pi}{4}$ is chosen to maximize the shift of the $k_y$-coordinate of the Dirac points in the strained region. As the strain $\epsilon$ increases from $-0.03$, the transmission angle corresponding to maximum transmission $T$ rotates along opposite directions in the two valleys. This occurs until a critical value of $\epsilon$ slightly above $0.023$ is reached, after which the maximum value of $T$ starts to decrease (see the transmission curve corresponding to $\epsilon = 0.03$ in Fig. 4) and eventually reaches nearly zero for all incidence angles as the strain is increased further. These trends can be explained by considering the overlap of the $q_y$ values spanned by the lead and central segment FSs. Taking the $K'$ valley shown in Fig. 4a as an example, the central segment FS at $\epsilon = -0.04$ (right panel) is well separated from the source FS (black circle at the center) so that there is no overlap between the range of $q_y$ values spanned by the two FS. The transmission is thus negligibly small. Increasing the strain causes the central segment FS to shift along the positive $q_x$ and $q_y$ axes until it begins to overlap with the lower half of the lead FS, resulting in significant transmission occurs at negative values of incidence angles. As the strain increases even further, the central segment FS eventually encloses the smaller lead FS and the incidence angle corresponding to peak transmission increases to $\varphi = 0$ (i.e. normal incidence) where the pseudospin orientations of the lead and central segment eigenstates match with one other. A small increase in the strain for which the source FS is still entirely enclosed within the central FS would cause the pseudospins of the source and central FS at $q_y = 0$ to become misaligned. The peak transmission therefore occurs not at normal incidence but at some positive value of the incidence angle at which the pseudospin orientations of the source and central FSs are aligned to one another. A further increase in the strain would cause the central segment FS to shift out of the lead FS, reducing the overlap between them and the overall transmission.

We note that the reciprocal space size of the central FSs decreases as the value of $\epsilon$ increases from a negative value to a positive value. This decrease in the size of the central FS is primarily because of the strain-induced shift in the energy of the Dirac point due to the $v_F w \boldsymbol{\sigma_0}$ term in Eq. (13). The energy shift is in turn proportional to the second-nearest neighbor coupling $t_n$ [Eq. (14)], which has been neglected in previous theoretical models of transmission through strained graphene heterostructures. The corresponding central FSs in the K valley where $t_n$ was set to 0 are plotted with dotted lines in the right plot of Fig. 4b, from which it can be seen that the resulting Fermi surfaces have a constant size that does not vary with the strain magnitude. Since the size of central FS directly affects its reciprocal space overlap with the source FS, and hence the resulting transmission profile, this demonstrates that the second-nearest neighbor coupling has to be taken into account to accurately model the transmission profiles.

In addition, the FSs are displaced in the opposite direction in $k$-space by the strain in the $K$ valley. Thus, at sufficiently large strain magnitude (of around $\epsilon = 0.018$), the transmission profiles corresponding to the two valleys will be well-separated with virtually no angular overlap in their transmission lobes. In this situation, one can place an insulating barrier in the heterojunction system to prevent transmission from the incidence angles that are predominantly transmitted by one valley (as we shall see later in Fig. 6). Thus the current that transmits through and gets collected at the drain would predominantly come from the other valley, resulting in high valley polarization [45] . In this manner, the device can then function as a strain valley filter for possible valleytronic applications [32, 33, 35, 37, 40 – 47] .

Another means of modulating the transmission profile is by varying the strain angle while keeping the central segment gate potential and strain magnitude fixed, as shown in Fig. 5. Similar to the effect of varying the strain magnitude, the variation of the strain angle causes a displacement of the central FS in reciprocal space relative to that of the source. At large negative values of the strain angle (at $\phi = -30^{\circ}$ say), the central FS does not overlap with the source FS and the resultant transmission becomes negligible. The maximum transmission magnitude of unity begins to occur when the angle of applied strain reaches a critical value (at $\phi \approx -20^{\circ}$ in the figure) at which the source and central FSs begins to overlap one other. Above the critical angle, the incidence angle corresponding to the maximum transmission increases (decreases) with the strain angle $\phi$ for the $K'$ ($K$) valley as the central segment FS moves into the source FS, then fully encloses it, and finally moves out of the source FS. The transmission eventually decreases almost to zero with further increase in the strain angle $\phi$ at which point the central segment FS lies separate from the lead FS with no overlap. This can be seen in Figs. 5a and 5b, where the transmission lobes of the $K'$ ($K$) valley almost disappear when the strain angle $\phi \geq 30^{\circ}$ ($\phi \leq -30^{\circ}$).

Finally, we calculate the net conductivity and the valley polarization through the strained graphene heterojunction, which are respectively given by the sum and the difference of the conductivity contributions from the two valleys ($K$ and $K'$). Since the Fermi surface of the unstrained lead is perfectly circular, the contribution to the net conductivity $\sigma$ at a given incidence angle $\varphi$ obeys the proportionality relation $d\sigma(\varphi) \propto (T_K + T_K') \cos \varphi$ where $T_{K/K'}$ denotes the transmission of the $K/K'$ valleys, while that of valley conductivity contribution obeys $d\sigma_V \propto (T_K - T_K') \cos \varphi$. Since $d\sigma_V$ is antisymmetric about $\varphi = 0$ as shown in Fig. 6d, a finite valley polarization can be achieved by imposing an asymmetry on the transmission process. This can be done, e.g., by incorporating an insulating barrier across the lower half of the heterojunction as shown in Fig. 6a so that the conductivity contributions for $\varphi < 0$ are blocked off and letting only the contributions from $\varphi \geq 0$ to be transmitted to the drain. In this case, the conductivity is obtained by integrating over the top half of the incidence angles, i.e., from $\varphi = 0$ to $\frac{\pi}{2}$, as $T = \int_0^{\frac{\pi}{2}} \mathrm{d}\sigma(\varphi)$. Similarly, we also calculate the valley polarization $P_V \equiv \left( \int_0^{\pi/2} (T_K - T'_K) \cos \varphi \ \mathrm{d}\varphi \right) / \left( \int_0^{\pi/2} (T_K + T'_K) \cos \varphi \ \mathrm{d}\varphi \right)$. The plots in Figs. 6b, 6c, and 6d respectively depict the effects of varying the gate voltage $V_0$, strain magnitude $\epsilon$ and strain angle $\phi$ on the conductivity and valley polarization. On the left of each figure, we plot the conductivity contributions of the individual valleys $K$ and $K'$ as a function of the angle of incidence $\varphi$, while on the right of each figure, we plot the total conductivity and valley polarization obtained after integration over all the positive incidence angles. From these plots, it can be observed that for all three parameters $V_0, \epsilon$ and $\phi$, there exist a range of values over which there is strong suppression of the conductance for both valleys, resulting in a nearly-zero OFF-state conductance. Hence, all three parameters can be utilized as externally controlled "knobs" to modulate or switch the electron transmission through the strained graphene heterostructure. In particular, one can achieve a strong suppression of the OFF current by varying the strain magnitude $\epsilon$ (see Fig. 6c) and achieve an ON-OFF conductance ratio of up to $10^{12}$. Similarly, there are also parameter ranges over which there is significant conductance in one valley whilst that of the other valley is suppressed to nearly zero. For instance, in the vicinity of $|V_0| = 0.4$ eV and $|\epsilon| = 0.025$, one obtains nearly conductance which is of almost perfect valley-polarization and having non-negligible magnitude, a combination which is useful for valleytronic applications.

## Conclusion

In conclusion, we have derived the full analytical expression for the Hamiltonian of graphene under a uniform uniaxial strain in the limit of small deformation and low energy. Our analysis shows that the physics of strained graphene cannot be fully captured by reducing the effect of strain to a single pseudo-gauge field, as was done in previous works. This is because in addition to the displacements of the Dirac cones in reciprocal space, the cones also undergo strain-induced rotation and tilt while the Dirac points are shifted in energy. We derived the coordinates of the Dirac points in $k$-space, and showed that the application of a uniaxial tensile or compressive strain shifts the Dirac points in the two valleys in opposite directions. This shift allows the design of a graphene-based heterojunction which can act as a current switch as well as generate valley-polarized current. We further derived the appropriate boundary conditions for the transmission across the interfaces in the heterojunction and the analytical expression for the transmission coefficient. We showed that the energy shift of the Dirac cones, which is proportional to the second-nearest coupling considered for the first time in a transport context in graphene here, should not be neglected for accurate transmission profile calculations. We also demonstrated that the transport properties of the heterojunction are valley asymmetric and can be tuned by modulating the gate voltage, and the strain magnitude and direction. An ON-OFF current ratio of up to $10^{12}$ as well as highly valley-polarized currents can potentially be achieved in the heterojunction by tuning the above parameters. The analytical and numerical results obtained provide a guide to the future realization of strain-modulated graphene devices.

**Acknowledgements**

This work is supported by the Ministry of Education (MOE) Tier-II grant MOE2018-T2-2-117 (NUS Grant Nos. R-263-000-E45-112/R-398-000-092-112), MOE Tier-I FRC grant (NUS Grant No. R-263-000-D66-114), and other MOE grants (NUS Grant Nos. C-261-000-207-532, and C-261-000-777-532).

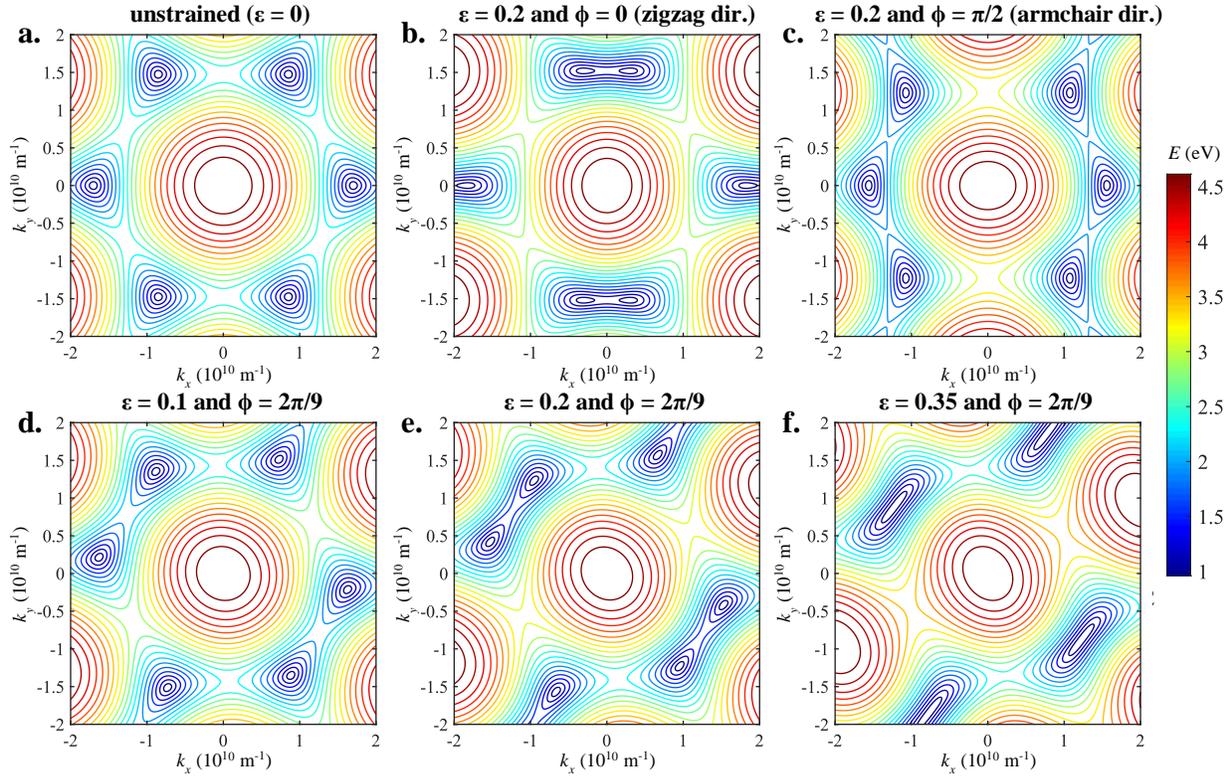

**Figure 1 | Contour plots of the conduction band.** Conduction band for various values of $\epsilon$ and $\phi$. The conduction band and the valence band touch at the Dirac points which are located at the darkest blue circles. (a) Unstrained graphene ($\epsilon = 0$). (b) Strained graphene in the zigzag direction with $\epsilon = 0.2$ and $\phi = 0$. (c) Strained graphene along the armchair direction with $\epsilon = 0.2$ and $\phi = \pi/2$. (d) Strained graphene with $\epsilon = 0.1$ and $\phi = 2\pi/9$. (e) Strained graphene with $\epsilon = 0.2$ and $\phi = 2\pi/9$. (f) Strained graphene with $\epsilon = 0.35$ and $\phi = 2\pi/9$ after gap opening. The two Dirac points have merged. The Dirac points around $k_y = 0$ are not shifted in the $k_y$-direction whenever the strain is applied along the armchair or the zigzag directions.

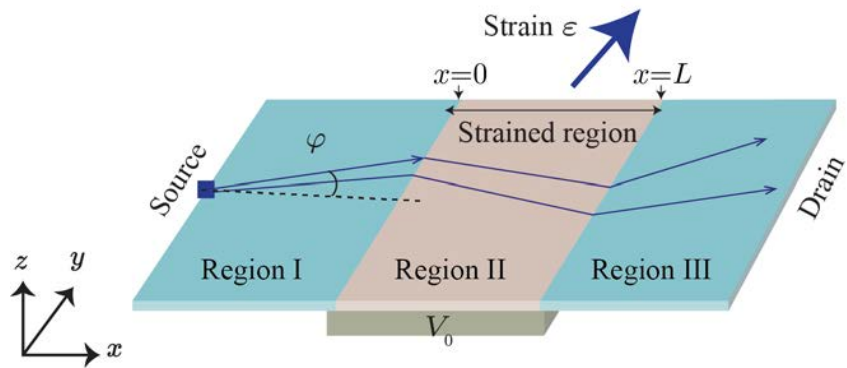

**Figure 2 | Schematic of the unstrained-strained-unstrained heterojunction graphene device.**

In the central region, a strain $\epsilon$ is applied along the $\phi$ direction and a gate potential of $V_0$ is applied.

The transmission is the proportion of the flux coming from the source that reaches the drain.

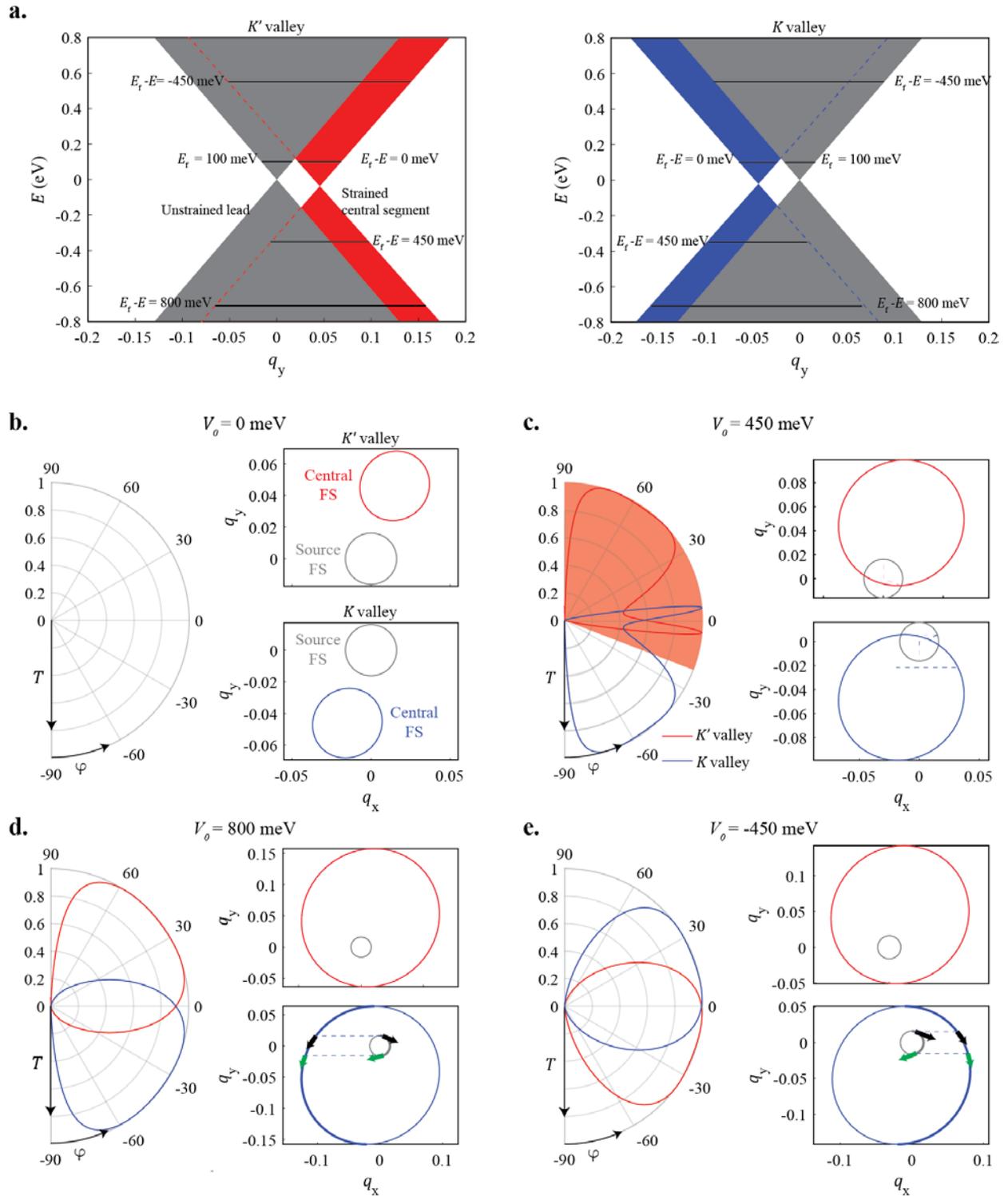

**Figure 3 | Impact of gate voltage on the transmission profile.** a. Projections of the dispersion

relations in the unstrained lead (gray) and strained central segment (blue/red) in the $K'$ (left) and $K$ (right) valleys on the $E - q_y$ plane when a strain of $\epsilon = 0.02$ is applied along $\phi = \pi/4$ at $E_{\mathrm{f}}$= 100 meV. The horizontal lines with the various values of $E_{\mathrm{f}} - E$ on the projections of the strained segment dispersion relations denote the sections which the Fermi surfaces in the strained segment lie on at the respective values of $V_0 = E_{\mathrm{f}} - E$ depicted in (b) – (e). (b) – (e) The left plots show the transmission coefficient $T = |t|^2$ as functions of the incident angle $\varphi$ when the incoming electron has an energy of 100 meV and passes through a potential barrier with $L =$ 15 nm at (b) $V_0$= 0 meV, (c) $V_0 = 450$ meV, (c) $V_0 = 800$ meV, and (d) $V_0 = $ -450 meV. The right plots show the Fermi surfaces of the leads and the central segments in the $K'$ and $K$ valleys. The angular range marked out in the source FS in (c) denotes the range of incidence angles at which the $q_y$ values of the source and central segments overlap. The corresponding range of overlapping incidence angles for the $K'$ valley is indicated by the light red background in the polar plot. The thicker portions of the $K$ valley FSs in (d) and (e) mark out the halves of the FS in which the states propagate in the positive $x$ direction. The arrows on the FSs denote the eigenspinor orientation of the eigenstates propagating in the positive $x$ direction at the same value of $q_y$ on the source and drain FSs

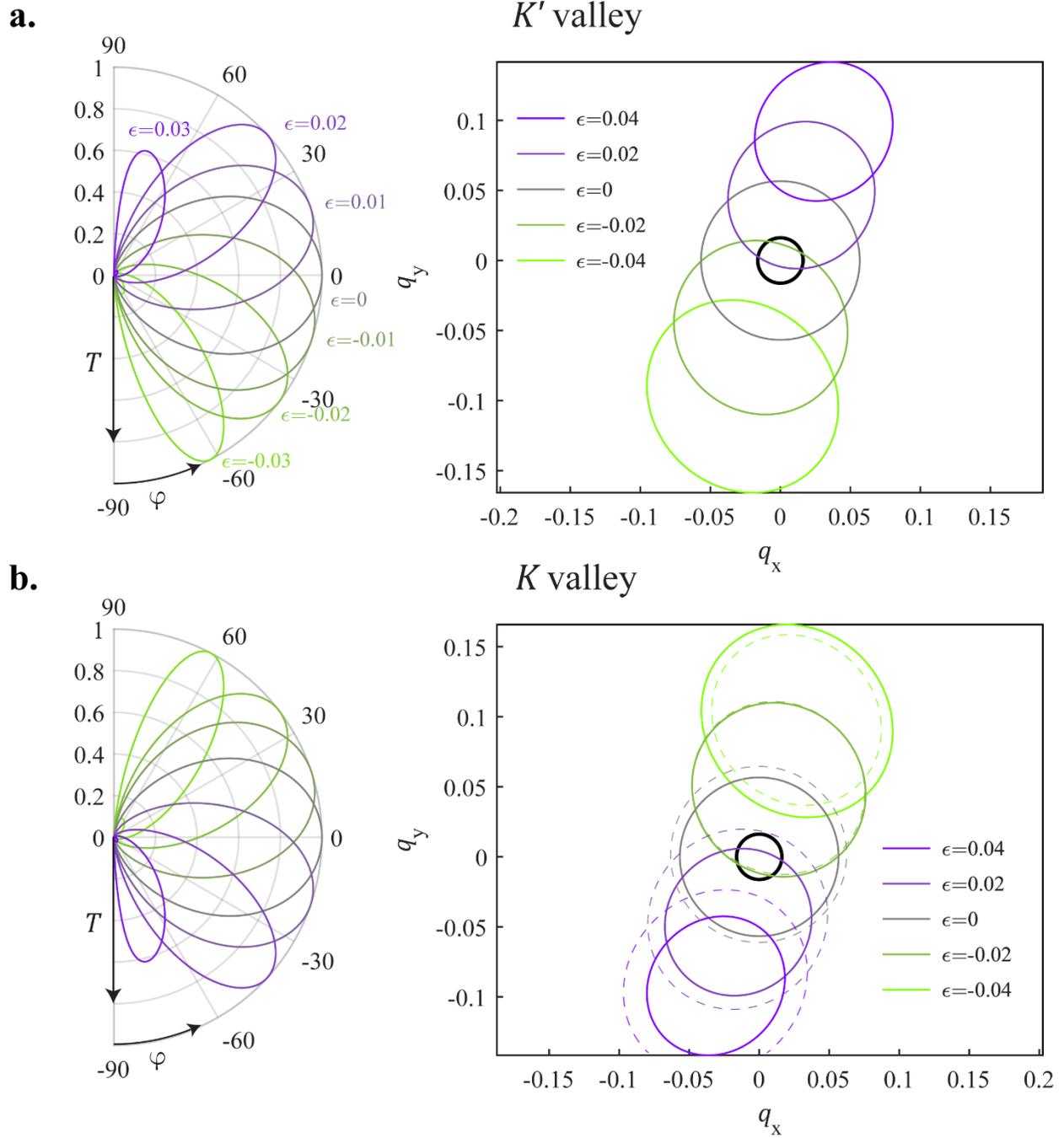

**Figure 4 | Impact of the strain magnitude on the transmission profile.** (left) Plots of the transmission coefficient $T = |t|^2$ as functions of the incident angle $\varphi$ for the (a) $K'$ and (b) $K$ valleys when the electron, which has an energy of 100 meV, passes through a central barrier segment of length $L = 4$ nm with an applied potential $V_0 = 0.45$ eV and strain at an angle of $\phi =$

$\pi/4$ and different magnitudes $\epsilon$ denoted by the line colours. The right plots show the source FSs (thick black line) and the central segment FSs at $\epsilon$ ranging from -0.04 to 0.04 denoted by the different colors in steps of 0.02. The dotted lines in the right plot of panel b shows the corresponding FSs of the central segments at the various values of $\epsilon$ when the next-nearest neighbour coupling $t_n$ is neglected.

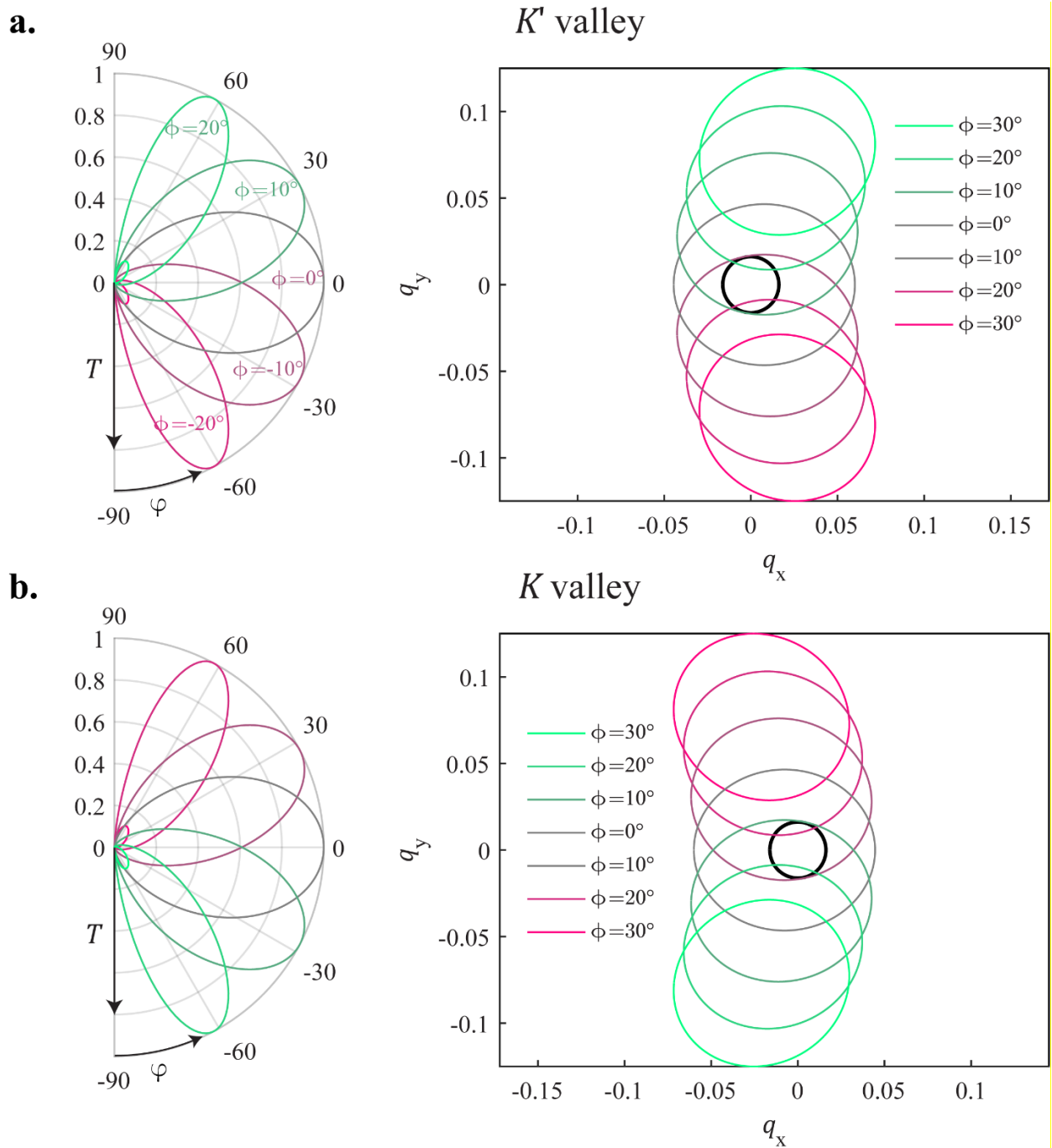

**Figure 5 | Impact of the strain angle on the transmission profile.** (left) Plots of the transmission coefficient $T = |t|^2$ as functions of the incidence angle $\varphi$ in the (a) $K'$ and (b) $K$ valleys when the electron has an energy of 100 meV passing through a central barrier segment of length $L = 4$ nm

with an applied potential $V_0 = 0.45\,\text{eV}$ and strain with a magnitude of $\epsilon = 0.04$ and different strain angles $\phi$ denoted by the line colors. The right plots show the source FSs (thick black line) and the central segment FSs at $\phi$ ranging from -30° to 30° denoted by the different colors in steps of 10°.

**a.**

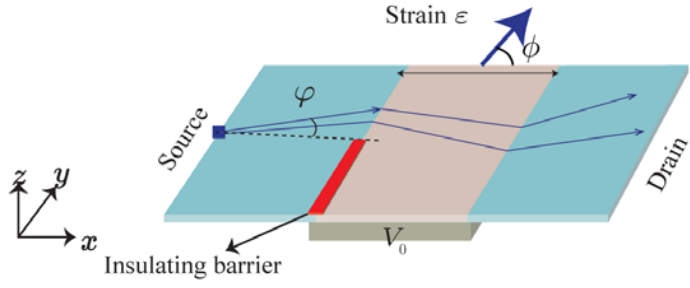

**b.**

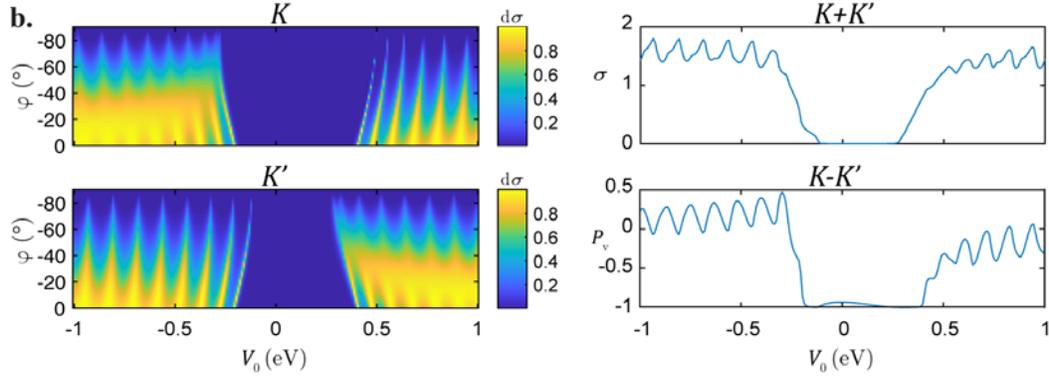

**c.**

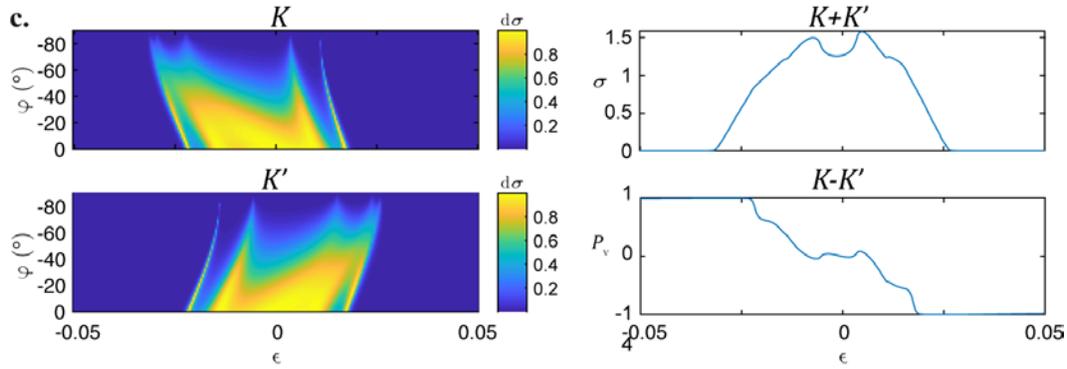

**d.**

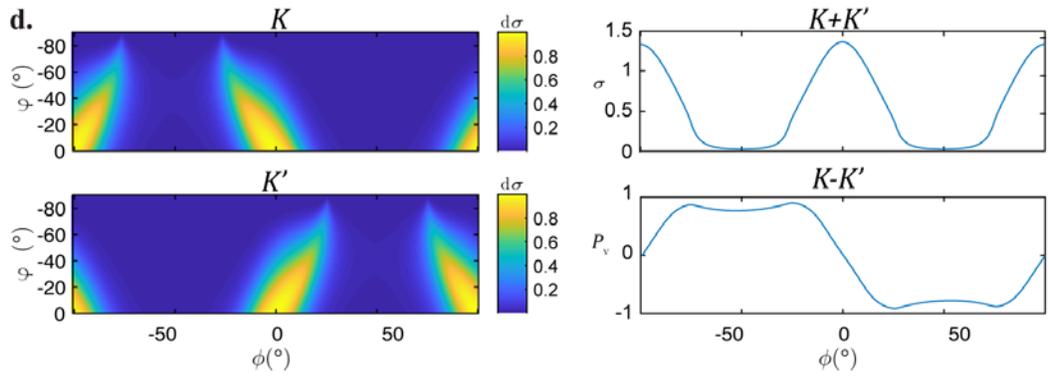

**Figure 6 | Charge conductivity and valley polarization through the strained graphene heterojunction.** (a) Modified strained heterojunction with an insulating barrier to block out transmission from incidence angles $\varphi < 0$. (b) Plots of the conductivity contributions from the $K$ valley (top left) and $K'$ valley (bottom left) for $\varphi \geq 0$, and the net conductivity (top right) and valley polarization (bottom right) as functions of the incidence angle $\varphi$ and (a) the gate voltage $V_0$, (b) strain magnitude $\epsilon$, and (c) strain angle $\phi$ for the parameters in Fig. 3, 4, and 5, respectively. The lower right plots show the net conductivity obtained over the incidence angles.